\documentclass[12pt]{article}

\newcommand{\be}{\begin{equation}}
\newcommand{\ee}{\end{equation}}
\newcommand{\bi}[1]{\vspace{-3mm} \bibitem{#1}}
\usepackage{epsfig,amsmath,amssymb,graphics,graphicx} 

\usepackage{graphics,graphicx}
\usepackage{amssymb,amsmath}
\usepackage{hyperref}
\usepackage{amsfonts}
\usepackage{epsfig}

\begin{document}

\begin{center}

{\it Modern Physics Letters B. Vol.28. No.7. (2014) 1450054} \\

\vskip 3mm

{\bf \large General Lattice Model of Gradient Elasticity} \\

\vskip 7mm
{\bf \large Vasily E. Tarasov} \\
\vskip 3mm

{\it Skobeltsyn Institute of Nuclear Physics,\\ 
Lomonosov Moscow State University, Moscow 119991, Russia} \\
{E-mail: tarasov@theory.sinp.msu.ru} \\

\begin{abstract}
New lattice model for the gradient elasticity is suggested.
This lattice model gives a microstructural basis 
for second-order strain-gradient elasticity of continuum 
that is described by the linear elastic constitutive relation
with the negative sign in front of the gradient. 
Moreover the suggested lattice model allows us to have 
a unified description of gradient models 
with positive and negative signs of the strain gradient terms.
Possible generalizations of this model for the high-order 
gradient elasticity and three-dimensional case 
are also suggested.
\end{abstract}

\end{center}

\noindent


\section{Introduction}

The two most widely used theories of elastic deformation 
in solid materials are a microscopic approach based on 
the statistical mechanics of lattices \cite{Born,MMW,Bo} and 
the quantum theory of solid-states \cite{Kittel},
and a macroscopic approach based on 
the classical continuum mechanics \cite{Sedov,Landau}.
Continuum elasticity is a phenomenological theory 
representing continuum limit of lattice dynamics,
where the length-scales are much larger than inter-atomic distances. 
Nonlocal elasticity theory is based on the assumption that 
the forces between material points can be at long-range 
in character, thus reflecting the long-range character 
of interatomic and intermolecular forces. 
In general, the nonlocal continuum models describe materials  
whose behavior at any point depends on the states of 
all other points in the media,
in addition to its own state and the state of external fields. 
Such considerations are well-known in solid-state physics, 
where the nonlocal interactions between the atoms and molecules
are prevalent in determining the properties of the media and materials. 

The theory of nonlocal continuum mechanics was formally initiated 
by the papers of Kroner \cite{Kroner} and 
Eringen, Edelen \cite{Eringer1972,EE}.
Kroner \cite{Kroner} indicated the relation between
nonlocal elasticity theory of  materials with long range cohesive forces.
Eringen and Edelen \cite{EE} provided derivation of 
the constitutive equations for the nonlocal elasticity. 
Eringen and Kim \cite{EK1977} described a relation between 
non-local elasticity and lattice dynamics. 
Kunin described the physical aspects of nonlocal elasticity 
in the book \cite{Kunin}, and studied various problems in Fourier space.
In the book \cite{Eringer2002} Eringen considered 
a unified approach to field theories for 
elastic solids, viscous fluids, and 
heat-conducting electromagnetic solids and fluids 
that include nonlocal effects. 
Rogula \cite{Rogula} considered the mathematical aspects of 
nonlocal elasticity models, 
proposed different types of nonlocal constitutive relations 
between stress and strain, and applied it to various problems 
in continuum mechanics. 
Non-local continuum mechanics has been treated with two different approaches \cite{Rogula,AA2011}:
the gradient elasticity theory (weak non-locality) and 
the integral non-local theory (strong non-locality).
In this paper we discuss the gradient models 
of non-locality elasticity.  
Usually two classes of gradient models are distinguished
by the different signs of the strain gradient terms
in the constitutive relations for the strain $\varepsilon_{ij}$ and the stress $\sigma_{ij}$:  
\be \label{H-1pm}
\sigma_{ij} = \Bigl( \lambda  \varepsilon_{kk} \delta_{ij} + 2 \mu \varepsilon_{ij} \Bigr)
\pm l^2 \, \Delta \, \Bigl( \lambda  \varepsilon_{kk} \delta_{ij} + 2 \mu \varepsilon_{ij} \Bigr) ,
\ee
where $\lambda$ and $\mu$ are the Lame coefficients, $l$ is the scale parameter.
If $l^2=0$, we have the classical case of the linear elastic constitutive 
relations for isotropic case that is the well-known Hooke's law.

The first class of gradient elasticity models 
are described by equations (\ref{H-1pm}) with the positive sign in front of gradient.
The main motivation to use this form of the gradient elasticity 
is the description of dispersive wave propagation 
through heterogeneous media. 
In many studies, gradient elasticity models 
with the positive sign in (\ref{H-1pm}) 
have been derived from associated lattice models by the 
continualization procedure for the response of a lattice 
\cite{M1968,RRG1995,MO1996}. 

The second class of gradient elasticity models 
are described by equations (\ref{H-1pm}) with the negative sign in front of the gradient.
The strain gradients in equation (\ref{H-1pm}) 
with the negative sign are equivalent to those derived 
from the positive-definite deformation energy density, 
and therefore these models of the strain gradients are stable. 

The positive sign of the strain gradient term 
in equation (\ref{H-1pm}) makes this term destabilizing.
The corresponding equation for the displacements 
is unstable for wave numbers $k> 1/l^2$ 
\cite{ASS2002,MA2002,RRG1995,GF2001}.
In dynamics the instabilities lead to an unbounded growth
of the response in time without external work. 
It is known the instabilities are related 
to loss of uniqueness in static boundary value problems. 
Instabilities in statics and dynamics for 
the second-gradient models with the positive sign 
are discussed in \cite{AM2002}. 

At this moment there is the opinion that 
gradient elasticity models with the negative sign in 
equations (\ref{H-1pm}) 
cannot be obtained from lattice models \cite{AA2011}.
It is usually assumed that this class of the second-gradient models does not 
have a direct relationship with discrete microstructure and 
lattice models  \cite{ASS2002}.
It was proved that the homogenization (continualization) procedure, 
which is considered in \cite{M1968,RRG1995,MO1996,MA2002,AM2005}, 
uniquely leads to a second-order
strain gradient term that is preceded by a positive sign.
The second-gradient model with negative sign
cannot be derived by this homogenization procedure.
From a mathematical point of view it is caused by properties
of the Taylor series that is used in this procedure.

In this article we propose lattice models,
that allow us to derive linear elastic constitutive relations 
with negative and positive signs.
Moreover the suggested lattice models give 
unified description of the gradient models 
with positive and negative signs of the strain gradient terms.
To obtain continuum equation from the lattice equations 
we use an approach that is suggested in
\cite{JMP2006,JPA2006,Chaos2006,TarasovSpringer}.


\section{Equations of lattice model}

Let us consider the vibration of an unbounded homogeneous lattice, 
such that all particles are displaced from its equilibrium position 
in one direction, and the displacement of particle 
is described by a scalar field. 
We consider one-dimensional lattice system of interacting particles,
where the equation of motion of $n$th particle is
\be \label{Main_Eq}
M \frac{d^2 u_n(t)}{d t^2} = 
g_2 \, \sum_{\substack{m=-\infty \\ m \ne n}}^{+\infty} \; K_2(n,m) \; u_m(t)
+ g_4 \, \sum_{\substack{m=-\infty \\ m \ne n}}^{+\infty} \; K_4(n,m) \; u_m(t)
+ F (u_n(t)) ,
\ee
where $u_n(t)$ are displacements from the equilibrium, $g_2$ and $g_4$ are coupling constants, 
$F(u_n)$ is the external on-site force,
$K_2(n,m)$ and $K_4(n,m)$ are the functions 
with different power-law asymptotic behavior of the functions
\be \label{Ksk-2}
\hat{K}_s(k) = 2 \sum^{\infty}_{n=1} K_s(n,0) \cos(kn) , \quad (s=2;4)
\ee
for $k \to 0$. We will consider interactions terms for 
which the difference $\hat{K}_s(k) - \hat{K}_s(0)$ 
are asymptotically equivalent to $|k|^s$ as $|k| \to 0$. 
Note some general properties of $K_s(n,m)$, with $s=2;4$. 
The conservation law of the total momentum in the lattice (\ref{Main_Eq})
in case of absence of external forces $F (u_n)=0$ gives
\be \label{Ko21}
\sum_{\substack{m=-\infty \\ m \ne n}}^{+\infty} \; K_2 (n,m) = 0 , \qquad
\sum_{\substack{m=-\infty \\ m \ne n}}^{+\infty} \; K_4 (n,m) = 0 , 
\ee
for all $n$. For homogeneous unbounded lattice, we have 
\[ K_2(n, m) = K_2(n - m) , \quad K_4(n, m) = K_4(n - m) , \]
where elements of $K_s(n, m)$ are constrained 
by the conditions (\ref{Ko21}), and
\be \label{Ko22}
\sum_{m} K_s(n-m) = \sum_{n} K_s(n-m) = 0 . \ee
For a simple case each particle is an inversion center and 
$K_s(n-m) = K_s(|n-m|)$, where $s=2;4$.
Using the condition (\ref{Ko21}), 
we can rewrite equation (\ref{Main_Eq}) as
\be \label{Main_Eq2}
M \frac{d^2 u_n}{d t^2} = 
g_2 \, \sum_{\substack{m=-\infty \\ m \ne n}}^{+\infty} \; 
K_2(|n-m|) \; \Bigl( u_n-u_m \Bigr) + 
g_4 \, \sum_{\substack{m=-\infty \\ m \ne n}}^{+\infty} \; 
K_4(|n-m|) \; \Bigl( u_n-u_m \Bigr) + F (u_n) .
\ee
In this form of equation of motion the interaction terms are translation invariant. 
It should be noted that the noninvariant terms lead to the divergences 
in the continuum models \cite{TarasovSpringer}.

Let us give an effective discrete mass-spring system
for the suggested lattice model (\ref{Main_Eq2}).
In Figure 1, we present the nearest-neighbor and 
next-nearest-neighbor interactions only. 
In general, the functions $K_2(|n-m|)$ and $K_4(|n-m|)$ 
describe long-range interactions with power-law asymptotic 
of (\ref{Ksk-2}).


\vskip -40mm
\begin{picture}(350,250)
\multiput(60,20)(80,0){5}{\circle{30}}
\multiput(75,20)(80,0){5}{\line(1,0){5}}
\multiput(45,20)(80,0){5}{\line(-1,0){5}}
\multiput(0,20)(80,0){6}{\line(1,-2){5}}
\multiput(35,10)(80,0){6}{\line(1,2){5}}
\multiput(5,10)(80,0){6}{\line(1,4){5}}
\multiput(10,30)(80,0){6}{\line(1,-4){5}}
\multiput(15,10)(80,0){6}{\line(1,4){5}}
\multiput(20,30)(80,0){6}{\line(1,-4){5}}
\multiput(25,10)(80,0){6}{\line(1,4){5}}
\multiput(30,30)(80,0){6}{\line(1,-4){5}}
\put(52,17){\text{n-2}}
\put(132,17){\text{n-1}}
\put(217,17){\text{n}}
\put(290,17){\text{n+1}}
\put(370,17){\text{n+2}}
\put(95,40){\text{$k^{eff}_2$}}
\put(175,40){\text{$k^{eff}_2$}}
\put(255,40){\text{$k^{eff}_2$}}
\put(335,40){\text{$k^{eff}_2$}}
\multiput(60,5)(160,0){3}{\line(-1,-1){35}}
\multiput(60,5)(160,0){3}{\line(+1,-1){35}}
\multiput(125,-40)(160,0){2}{\line(1,4){5}}
\multiput(130,-20)(160,0){2}{\line(1,-4){5}}
\multiput(135,-40)(160,0){2}{\line(1,4){5}}
\multiput(140,-20)(160,0){2}{\line(1,-4){5}}
\multiput(145,-40)(160,0){2}{\line(1,4){5}}
\multiput(150,-20)(160,0){2}{\line(1,-4){5}}
\multiput(0,-30)(160,0){3}{\line(1,0){25}}
\multiput(95,-30)(160,0){3}{\line(1,0){25}}
\multiput(120,-30)(160,0){2}{\line(1,-2){5}}
\multiput(155,-40)(160,0){2}{\line(1,2){5}}
\put(135,-15){\text{$k^{eff}_4$}}
\put(295,-15){\text{$k^{eff}_4$}}
\multiput(140,35)(160,0){2}{\line(1,1){35}}
\multiput(140,35)(160,0){2}{\line(-1,1){35}}
\multiput(45,60)(160,0){3}{\line(1,4){5}}
\multiput(50,80)(160,0){3}{\line(1,-4){5}}
\multiput(55,60)(160,0){3}{\line(1,4){5}}
\multiput(60,80)(160,0){3}{\line(1,-4){5}}
\multiput(65,60)(160,0){3}{\line(1,4){5}}
\multiput(70,80)(160,0){3}{\line(1,-4){5}}
\multiput(80,70)(160,0){3}{\line(1,0){25}}
\multiput(15,70)(160,0){3}{\line(1,0){25}}
\put(0,70){\line(1,0){20}}
\put(425,70){\line(1,0){20}}
\multiput(40,70)(160,0){3}{\line(1,-2){5}}
\multiput(75,60)(160,0){3}{\line(1,2){5}}
\put(55,90){\text{$k^{eff}_4$}}
\put(215,90){\text{$k^{eff}_4$}}
\put(375,90){\text{$k^{eff}_4$}}
\put(55,43){\text{M}}
\put(135,43){\text{M}}
\put(215,43){\text{M}}
\put(295,43){\text{M}}
\put(375,43){\text{M}}
\multiput(60,5)(160,0){3}{\line(0,-1){80}}
\put(140,-65){\vector(1,0){80}}
\put(140,-65){\vector(-1,0){80}}
\put(300,-65){\vector(1,0){80}}
\put(300,-65){\vector(-1,0){80}}
\put(135,-60){\text{$a$}}
\put(295,-60){\text{$a$}}
\put(-20,-100){\text{Figure 1: Discrete mass-spring system
with effective stiffness coefficients $k^{eff}_2 =k^{eff}_2(g_2,g_4)$}}
\put(-20,-115){\text{and $k^{eff}_4 =k^{eff}_4(g_2,g_4)$, 
the mass $M$ and the distance $a$ that correspond to 
the lattice model }} 
\put(-20,-130){\text{with coupling constants $g_2$ and $g_4$.}}
\end{picture}
\vskip 50mm

\section{From lattice model to continuum model}

Let us consider a set of operations  
\cite{JMP2006,JPA2006,TarasovSpringer} that transforms
the equations of motion of the lattice model 
into a continuum equation for the displacement field $u(x,t)$. 
We assume that $u_n(t)$ are Fourier coefficients
of the field $\hat{u}(k,t)$ on $[-K_0 /2, K_0 /2]$ 
that is described by the equations
\be \label{ukt}
\hat{u}(k,t) = \sum_{n=-\infty}^{+\infty} \; u_n(t) \; e^{-i k x_n} =
{\cal F}_{\Delta} \{u_n(t)\} ,
\ee
\be \label{un} 
u_n(t) = \frac{1}{K_0} \int_{-K_0/2}^{+K_0/2} dk \ \hat{u}(k,t) \; e^{i k x_n}= 
{\cal F}^{-1}_{\Delta} \{ \hat{u}(k,t) \} ,
\ee
where $x_n = n \, a $ and $a =2\pi/K_0$ is 
distance between equilibrium positions of the lattice particles. 
Equations (\ref{ukt}) and (\ref{un}) are the basis for 
the Fourier transform ${\cal F}_{\Delta}$ and the inverse 
Fourier series transform ${\cal F}^{-1}_{\Delta}$.
The Fourier transform can be derived from (\ref{ukt}) and (\ref{un}) 
in the limit as $a  \to 0$ ($K_0 \to \infty$). 
In this limit ($a \to 0$ or $K_0 \to \infty$) the sum 
becomes the integral, and 
equations (\ref{ukt}) and (\ref{un}) become
\be \label{ukt2} 
\tilde{u}(k,t)=\int^{+\infty}_{-\infty} dx \ e^{-ikx} u(x,t) = 
{\cal F} \{ u(x,t) \}, 
\ee
\be \label{uxt}
u(x,t)=\frac{1}{2\pi} \int^{+\infty}_{-\infty} dk \ e^{ikx} \tilde{u}(k,t) =
 {\cal F}^{-1} \{ \tilde{u}(k,t) \} . 
\ee
Here we use the lattice function
\[ u_n(t) = \frac{2 \pi}{K_0} u(x_n,t) \] 
with continuous $u(x,t)$, where 
\[ x_n=n \, a = \frac{2 \pi n}{K_0} \to x . \]
We assume that $\tilde{u}(k,t)= {\cal L} \hat{u}(k,t)$,
where ${\cal L}$ denotes the passage 
to the limit $a \to 0$ ($K_0 \to \infty$),
i.e. the function $\tilde{u}(k,t)$ can be derived 
from $\hat{u}(k,t)$ in the limit $a \to 0$.
Note that $\tilde{u}(k,t)$ is a Fourier transform of the field $u(x,t)$.
The function $\hat{u}(k,t)$ is a Fourier series transform of $u_n(t)$,
where we can use $u_n(t)=(2\pi/K_0) \, u(n \, a ,t)$.

We can state that a lattice model transforms 
into a continuum model 
by the combination ${\cal F}^{-1} {\cal L} \ {\cal F}_{\Delta}$ 
of the following operation. \\
The Fourier series transform:
\be \label{O1}
{\cal F}_{\Delta}: \quad u_n(t) \to {\cal F}_{\Delta}\{ u_n(t)\}=
\hat{u}(k,t) .
\ee
The passage to the limit $a \to 0$:
\be \label{O2}
{\cal L}: \quad \hat{u}(k,t) \to {\cal L} \{\hat{u}(k,t)\}=
\tilde{u}(k,t) . \ee
The inverse Fourier transform: 
\be \label{O3}
{\cal F}^{-1}: \quad \tilde{u}(k,t) \to 
{\cal F}^{-1} \{ \tilde{u}(k,t)\}=u(x,t) .
\ee
These operations allow us to get a continuum model
from the lattice model \cite{JMP2006,JPA2006,TarasovSpringer}.

\section{Lattice model with nearest-neighbor interaction}

Let us derive the usual elastic equation 
from the lattice model with the nearest-neighbor interaction
with coupling constant $g_2=K$ by the method suggested in 
\cite{JMP2006,JPA2006,TarasovSpringer}.
We will use equations (\ref{Main_Eq}) with
\be
\sum_{\substack{m=-\infty \\ m \ne n}}^{+\infty} \; K_2(n,m) \; u_m(t) =
u_{n+1}(t)-2u_n(t)+u_{n-1}(t) , \quad  K_4(n,m)=0 ,
\ee
where the term $K_2(n,m)$ describes the nearest-neighbor interaction.

We can give the following statement regarding the lattice model with the nearest-neighbor interaction 
and the corresponding continuum equation 
that is obtained in the limit $a \to 0$.

{\bf Proposition 1.} 
{\it In the continuous limit ($a \to 0$) the lattice equations of motion 
\be \label{CEM}
M \frac{d^2 u_n(t)}{d t^2} = 
K \cdot \Bigl( u_{n+1}(t)-2u_n(t)+u_{n-1}(t) \Bigr) + F (u_n(t))
\ee
are transformed by the combination ${\cal F}^{-1} {\cal L} \ {\cal F}_{\Delta}$ 
of the operations (\ref{O1}-\ref{O3}) into the continuum equation: 
\be \label{CME0}
\frac{\partial^2 u(x,t)}{\partial t^2} = C^2_e \, \frac{\partial^2 u(x,t)}{\partial x^2} + \frac{1}{\rho} f(u) ,
\ee
where $C^2_e = E / \rho ={K \, a^2}/M$ is a finite parameter, and $f(u) = F(u)/ (A \, a)$.}

{\bf Proof}.
To derive the equation for the field $\hat u(k,t)$, we
multiply equation (\ref{CEM}) by $\exp(-ik \, n \, a )$, 
and summing over $n$ from $-\infty$ to $+\infty$. Then
\be \label{DD1}
M \, \sum^{+\infty}_{n=-\infty} e^{-ik \,n \, a } \frac{d^2 u_n}{d t^2} 
=
K \cdot \sum^{+\infty}_{n=-\infty} \, e^{-ik \, n \, a} 
\, \Bigl( u_{n+1}-2u_n+u_{n-1} \Bigr) +
\sum^{+\infty}_{n=-\infty} e^{-ik \, n \, a} F(u_n) .
\ee
The first term on the right-hand side of (\ref{DD1}) is
\[
K \cdot \sum^{+\infty}_{n=-\infty} \
e^{-ik \, n \, a} K_2(n,m) u_n = K \cdot \sum^{+\infty}_{n=-\infty} \
e^{-ik \, n \, a} \, \Bigl( u_{n+1}-2u_n+u_{n-1} \Bigr) 
= \]
\[ = K \cdot \sum^{+\infty}_{n=-\infty} \
e^{-ikn d }  u_{n+1} -
2 K \cdot \sum^{+\infty}_{n=-\infty} \
e^{-ikn d }  u_n +
K \cdot \sum^{+\infty}_{n=-\infty} \
e^{-ik \, n \, a}  u_{n-1}= \]
\[ =e^{ik \, a } \, 
K \cdot \sum^{+\infty}_{m=-\infty} \
e^{-ik \, m \, a }  u_{m} -
2 K \cdot \sum^{+\infty}_{n=-\infty} \
e^{-ik\, n \, a }  u_n + e^{-ik d } 
K \cdot \sum^{+\infty}_{j=-\infty} \
e^{-ik \, j \, a }  u_{j} .
\]
Using the definition of $\hat{u}(k,t)$, we obtain
\[ K \cdot \sum^{+\infty}_{n=-\infty} \
e^{-ik \, n \, a } K_2(n,m) u_n=
K \cdot \Bigl( e^{ikd } \hat{u}(k,t)- 2 \hat{u}(k,t) +
e^{-ik \, a } \hat{u}(k,t) \Bigr)= \]
\be \label{Proof-nearest1}
= K \cdot \Bigl( e^{ik \, a } +e^{-ik \, a} -2 \Bigr) \hat{u}(k,t)= 
2 K \cdot \Bigl( \cos \left( k \, a \right)-1 \Bigr) \, \hat{u}(k,t) 
= -4 K \cdot \sin^2 \left( \frac{k\, a}{2} \right) \hat{u}(k,t) .
\ee
Substitution of (\ref{Proof-nearest1}) into (\ref{DD1}) gives
\be \label{simple}
M \frac{\partial^2 \hat{u}(k,t)}{\partial t^2} = 
-4 K \cdot \sin^2 \left( \frac{k\, a}{2}  \right) \hat{u}(k,t) +
\mathcal{F}_{\Delta} \{ F \left( u_n(t) \right) \} .
\ee
For $a \to 0$, the asymptotic behavior of the sine is
$\sin( k \, a /2 ) = k \, a/2 + O((k\, a)^3)$, 
then 
\[  - 4 \, \sin^2 \left( \frac{k\, a}{2}  \right) 
= - \left( k \, a \right)^2 + O((k\, a)^4) . \]
Using the finite parameter $C^2_e=K \, a^2/M$, 
the transition to the limit $a \to 0$ 
in equation (\ref{simple}) gives 
\be \label{DD2}
\frac{\partial^2  \tilde u(k,t)}{\partial t^2}=
- C^2_e k^2 \tilde u(k,t) +  \frac{1}{M} {\cal F} \{F(u)\} ,
\ee
where 
\be
\rho=\frac{M}{A \, a} ,\qquad E= \frac{K \, a}{A}, 
\qquad C^2_e =\frac{E}{\rho}=\frac{K \, a^2}{M} .
\ee
The inverse Fourier transform ${\cal F}^{-1}$ of (\ref{DD2}) has the form
\[ \frac{\partial^2 {\cal F}^{-1}\{ \tilde u(k,t)\} }{\partial t^2}=
- C^2_e {\cal F}^{-1} \{k^2 \tilde u(k,t)\} + \frac{1}{\rho} f(u) , \]
where $f(u)=F(u)/(A \, a)$ is the force density. 
Then we can use ${\cal F}^{-1}\{ \tilde u(k,t)\} =\tilde u(x,t)$,
and the connection between derivatives and its Fourier transform: 
${\cal F}^{-1} \{k^2 \tilde u(k,t)\} =\partial^2 u(x,t) / \partial x^2$. 
As a result, we obtain 
the continuum equation (\ref{CME0}). This ends the proof. \\

As a result, we prove that the lattice equations (\ref{CEM}) 
in the limit $a \to 0$ give the continuum equation 
with derivatives of second order only.
This conclusion agrees with the results of \cite{Maslov},
where the relation
\[ \exp \, i \left( -i \, a \, \frac{\partial}{\partial x} \right) \, u(x,t) = u(x+a,t) \]
and the representation of (\ref{CEM}) by pseudo-differential equation are used.


\section{From general lattice model to gradient elasticity model}

Let us consider the lattice model that is described by (\ref{Main_Eq}), 
where the terms $K_s(n,m)$ with $s=2$ and $s=4$ satisfy the conditions
\be \label{Jnm}
K_s(n,m)=K_s(|n-m|) , \qquad \sum^{\infty}_{n=1} |K_s(n)|^2 < \infty .
\ee
To describe gradient elasticity models, we consider the 
inter-particle interactions, that are described by $K_s(n)$ 
($s=2$ or $s=4$) of the following special type. 
We assume that the function
\be \label{Jak}
\hat{K}_s(k)=\sum^{+\infty}_{\substack{n=-\infty \\ n\not=0}} 
e^{-ikn} K_s(n) = 2 \sum^{\infty}_{n=1} K_s(n) \cos(kn) ,
\ee
satisfies the condition
\be \label{Aa}
\lim_{k \rightarrow 0} 
\frac{\hat{K}_s(k) - \hat{K}_s(0)}{|k|^s} = A_s ,
\ee
where $0<|A_s|< \infty$. Condition (\ref{Aa}) means that 
\be \label{AR}
\hat{K}_s(k)- \hat{K}_s(0) = A_s |k|^s +R_{s}(k),
\ee
for $k\rightarrow 0$, where
$ \lim_{k \rightarrow 0} \ R_s(k) / |k|^s  =0$.
This also means that, we can consider arbitrary 
functions $K_s(|n-m|)$ for which $\hat{K}_s(k)- \hat{K}_s(0)$ 
are asymptotically equivalent to $|k|^s$ as $|k| \to 0$.
 
As an example of the interaction terms $K_s(|n-m|)$, which 
give the continuum equations of gradient elasticity models, 
we consider the function
\be \label{Kn}
K_s(|n-m|) =
\frac{(-1)^{|n-m|}}{2\, \Gamma(s/2+1+|n-m|) \, \Gamma(s/2+1-|n-m|)} .
\ee
We use $2$ in the denominator to cancel with $2$ 
from equations (\ref{Jak}).
The terms $K_s(|n-m|)$ are considered for $n \ne m$, 
i.e. $|n-m| \ne 0$. 
For $s=2j$, we have $K_s(|n-m|)= 0$ for all $|n-m| \ge j+1$.
The function $K_s(n-m)$ with even value of $s=2j$ describes 
an interaction of the $n$-particle 
with $2 \,j$ particles with numbers 
$n \pm 1$ . . . $n \pm j$.
To represent properties of (\ref{Kn}), 
we can consider the function
\be \label{Knf}
f_K(x,y) = \operatorname{Re}[K_y(x)]
=\frac{ \operatorname{Re}[(-1)^{|x|}] }{2 \, \Gamma(y/2+1+|x|) \, \Gamma(y/2+1-|x|)} 
\ee
of two continuous variables $x$ and $y>0$.
Note that $\operatorname{Re}[(-1)^{|x|}] = (-1)^{|x|}$ 
for integer $x=n-m$.
The plots of the function (\ref{Knf}) are presented 
by Figures 2 and 3 for different ranges of $x$ and $y$.
This function decays rapidly with growth $x$ and $y$.
The function (\ref{Knf}) defines the interaction terms 
$K_s(|n-m|)$ by the equation $K_s(|n-m|)= f_K(|n-m|,s)$.

Using an inverse relation to (\ref{Jak}) with 
$\hat{K}_s(k) = |k|^s $ that has the form
\[ K_s(n) = \frac{1}{\pi} \int^{\pi}_0 k^s \, \cos(n \, k) \, dk   \]
we get another example of $K_s(|n-m|)$ in the form 
\be \label{Kn2}
K_s(|n-m|) = \frac{\pi^s}{s+1} \, _1F_2 \left(\frac{s+1}{2};
\frac{1}{2},\frac{s+3}{2};-\frac{\pi^2\, (n-m)^2}{4} \right) ,
\ee
where $\, _1F_2$ is the Gauss hypergeometric function 
(see Chapter II in \cite{Erdelyi}). 
Note that the interactions with (\ref{Kn2}) for $s=2$ and $s=4$ 
are long-range interactions of $n$-particle 
with all other particles ($m \in \mathbb{N}$).
It is easy to see that expression (\ref{Kn2}) is more complicated than (\ref{Kn}).

For $s=2$, we can also use the long-range interactions 
in the following two forms
\be
K_2(|n-m|) = \frac{(-1)^{|n-m|}}{(n-m)^2} , \quad
K_2(|n-m|) = \frac{1}{|n-m|^{\alpha}} , \quad (\alpha \ge 3) .
\ee

A main advantage of the interaction 
in the forms (\ref{Kn}) and (\ref{Kn2})
is a possibility to use for other generalizations
for the case of the high-order gradient elasticity 
by using arbitrary integer values of $s$
and the fractional generalization of gradient elasticity 
by non-integer values of $s$.


{\bf Proposition 2.}
{\it The lattice equations 
\be \label{C1}
M \frac{d^2 u_n}{d t^2} = 
g_2 \sum_{\substack{m=-\infty \\ m \ne n}}^{+\infty} 
\, K_2(|n-m|) \; [u_n -u_m] +
g_4 \sum_{\substack{m=-\infty \\ m \ne n}}^{+\infty} 
\, K_4(|n-m|) \; [u_n -u_m] + F (u_n) ,
\ee 
where $g_2$ and $g_4$ are coupling constants, 
$K_2(|n-m|)$ and $K_4(|n-m|)$ are defined by (\ref{Kn}), 
$u_n=u_n(t)$, 
are transformed by the combination ${\cal F}^{-1} {\cal L} \ {\cal F}_{\Delta}$ 
of the operations (\ref{O1}-\ref{O3}) into the continuum equation: 
\be \label{CME}
\frac{\partial^2 u(x,t)}{\partial t^2} -
G_2 \frac{\partial^2 u(x,t)}{\partial x^2 } + 
G_4 \frac{\partial^4 u(x,t)}{\partial x^4 } -
\frac{1}{\rho} f\left( u(x,t) \right) = 0  ,
\ee
where  
\be \label{G2G4}
G_2=\frac{g_2 \, a^2}{4 M} , \qquad G_4= \frac{g_4 a^4}{48 M} 
\ee
are finite parameters, $\rho=M/(A \, a)$ is the mass density
and $f(u)=F(u)/(A\, a)$ is the force density. } \\


{\bf Proof.}
To derive the equation for the field $\hat u(k,t)$, we
multiply equation (\ref{C1}) by $\exp(-ik \, n \, a)$, 
and summing over $n$ from $-\infty$ to $+\infty$. Then
\be \label{C3a}
M \, \sum^{+\infty}_{n=-\infty} e^{-ik \, n \, a} 
\frac{d^2}{d t^2}u_n(t)=  \sum^{+\infty}_{n=-\infty} \
\sum^{+\infty}_{\substack{m=-\infty \\ m \not=n}}
\sum_{s=2;4} e^{-ik \,n \, a} g_s \,  K_s(|n-m|) \ [u_n-u_m] +
\sum^{+\infty}_{n=-\infty} e^{-ik \, n \, a} F(u_n) .
\ee
The left-hand side of (\ref{C3a}) gives
\be
\sum^{+\infty}_{n=-\infty} e^{-ik \, n \, a} 
\frac{\partial^2 u_n(t)}{\partial t^2}=
\frac{\partial^2 }{\partial t^2}
\sum^{+\infty}_{n=-\infty} e^{-ik \, n \, a} u_n(t)=
\frac{\partial^2 \hat{u}(k,t)}{\partial t^2} ,
\ee
where $\hat{u}(k,t)$ is defined by (\ref{ukt}).
The second term of the right-hand side of equation (\ref{C3a}) is
$\sum^{+\infty}_{n=-\infty} e^{-ik \, n \, a} F(u_n)=
{\cal F}_{\Delta} \{F(u_n)\}$. 
The first term on the right-hand side of (\ref{C3a}) is
\[
\sum^{+\infty}_{n=-\infty} \ \sum^{+\infty}_{\substack{m=-\infty \\ m \not=n}} 
e^{-ik \, n \, a} K_s(|n-m|) \, [u_n-u_m] = \]
\be \label{C6}
=\sum^{+\infty}_{n=-\infty} \  \sum^{+\infty}_{\substack{m=-\infty \\ m \not=n}}
e^{-ik \, n \, a} K_s(|n-m|) \, u_n - 
\sum^{+\infty}_{n=-\infty} 
\sum^{+\infty}_{\substack{m=-\infty \\ m \not=n}} 
e^{-ik \, n \, a} K_s(|n-m|) \, u_m .
\ee
Using (\ref{ukt}) and (\ref{Jnm}), 
the first term in r.h.s. of (\ref{C6}) gives
\be \label{C7} 
\sum^{+\infty}_{n=-\infty} \ \sum^{+\infty}_{\substack{m=-\infty \\ m \not=n}}
e^{-ik \, n \, a} K_s(|n-m|) \, u_n =
\sum^{+\infty}_{n=-\infty} e^{-ik \, n \, a} u_n 
\sum^{+\infty}_{\substack{m^{\prime}=-\infty \\ m^{\prime} \not=0}}
K_s(m^{\prime})= \hat u(k,t) \hat{K}_s (0) ,
\ee
where 
\be \label{not}
\hat{K}_s(k \, a)=
\sum^{+\infty}_{\substack{n=-\infty \\ n\not=0}} 
e^{-ik \, n \, a} K_s(n)={\cal F}_{\Delta}\{ K_s(n)\} .
\ee
The second term in r.h.s. of (\ref{C6}) gives
\[ \sum^{+\infty}_{n=-\infty} \ 
\sum^{+\infty}_{\substack{m=-\infty \\ m \not=n}}
e^{-ik \, n \, a} K_s(|n-m|) u_m = 
\sum^{+\infty}_{\substack{n=-\infty \\ n \not=m}} 
e^{-ik \, n \, a} K_s(|n-m|) \ \sum^{+\infty}_{m=-\infty} u_m = \]
\be \label{C9}
= \sum^{+\infty}_{\substack{n^{\prime}=-\infty \\ n^{\prime}\not=0}} 
e^{-ik \, n^{\prime} \, a} K_s(n^{\prime}) \
\sum^{+\infty}_{m=-\infty } u_m e^{-ik \, m \, a}=
\hat{K}_s (k \, a) \ \hat u(k,t) .
\ee

As a result, equation (\ref{C3a}) has the form
\be \label{20}
M \frac{\partial^2  \hat u(k,t)}{\partial t^2}= \sum_{s=2;4}
\Bigl( \hat{K}_s (0)- \hat{K}_s (k \, a) \Bigr) \, \hat u(k,t) 
+{\cal F}_{\Delta} \{F(u_n)\} ,
\ee 
where ${\cal F}_{\Delta} \{F(u_n)\}$ is an operator notation for the Fourier
series transform of $F(u_n)$. 


Using the series (see Sec.5.4.8.12 in \cite{Prudnikov}) 
of the form
\[ \sum^{\infty}_{n=1}
\frac{(-1)^n}{\Gamma(\nu +1+n) \Gamma(\nu +1-n)} \cos(nk)
= \frac{2^{2 \nu -1}}{\Gamma(2 \nu +1)} 
\sin^{2\nu} \left(\frac{k}{2}\right) -\frac{1}{2\Gamma^2(\nu+1)} , \]
where $\nu >-1/2$ and $0<k<2\pi$, we get for 
the function (\ref{Jak}) of the form (\ref{Kn}) the equation
\be \label{Kk-K0}
\hat{K}_{s}(a \, k)-\hat{K}_{s}(0)= 
\frac{2^{s-1}}{\Gamma(s+1)} 
\sin^{s} \left(\frac{a\, k}{2} \right) = 
\frac{1}{2\Gamma(s+1)} \, |a\, k|^s + O(k^{s+2}) . 
\ee
Here we use $\nu=s/2$ and $ \sin (k/2) =k/2+ O(k^3)$. 
Note that $2$ in the denominator of (\ref{Kn}) cancels with $2$ 
from equation (\ref{Jak}) in front of the sum from zero 
to infinity. The limit $k \to 0$ gives
\be \label{kka}
\lim_{k \to 0} \frac{\hat{K}_{s}(k)- \hat{K}_{s}(0)}{|k|^{s}} = 
\frac{1}{2 \Gamma(s+1)} ,
\ee
and we have $A_s = 1  / (2 \Gamma (s+1))$. 


The Fourier series transform ${\cal F}_{\Delta}$ of (\ref{C1})
gives (\ref{20}).
We will be interested in the limit $a \rightarrow 0$. 
Using (\ref{Kk-K0}), equation (\ref{20}) can be written as
\be \label{Eq-k}
\frac{\partial^2}{\partial t^2} \hat{u}(k,t) -  
\frac{g_2 \, a^2}{M}  \; \hat{\mathcal{T}}_{2, \Delta}(k) \; \hat{u}(k,t) -
\frac{g_4 a^4}{M}  \; \hat{\mathcal{T}}_{4, \Delta}(k) \; \hat{u}(k,t)  
- \frac{1}{M} \mathcal{F}_{\Delta} \{ F \left( u_n(t) \right) \}  = 0, 
\ee
where 
\be 
\hat{\mathcal{T}}_{s, \Delta}(k) = 
- \frac{1}{2\Gamma(s+1)} |k|^s  + a^2 \, O(|k|^{s+2}) .
\ee

In the limit $a \rightarrow 0$, using
\be 
\hat{\mathcal{T}}_{s}(k) =
{\cal L}\hat{\mathcal{T}}_{s, \Delta}(k) = - \frac{1}{2 \Gamma(s+1)} |k|^{s} 
\quad (s=2;4) ,
\ee
we get
\be \label{T2T4}
\hat{\mathcal{T}}_2(k) = {\cal L}\hat{\mathcal{T}}_{2, \Delta}(k) = - \frac{1}{4} |k|^2 , \quad 
\hat{\mathcal{T}}_4(k) = {\cal L}\hat{\mathcal{T}}_{4, \Delta}(k) = - \frac{1}{48} |k|^4 .
\ee
The passage to the limit $a \rightarrow 0$ 
for the third term of (\ref{Eq-k}) gives
${\cal F}_{\Delta} F(u_n) \rightarrow 
{\cal L} {\cal F}_{\Delta} F(u_n)$.
Then
\be
{\cal L} {\cal F}_{\Delta} \{F(u_n)\} ={\cal F} \{ {\cal L} F(u_n)\} =
{\cal F} \{ F({\cal L} u_n)\}={\cal F} \{F(u(x,t))\} ,
\ee
where we use ${\cal L} {\cal F}_{\Delta} ={\cal F} {\cal L}$. 

As a result, equation (\ref{Eq-k}) in the limit $a \rightarrow 0$ gives
\be \label{Eq-k2}
\frac{\partial^2}{\partial t^2} \tilde{u}(k,t) - 
G_2 \; \hat{\mathcal{T}}_2(k) \; \tilde{u}(k,t) - 
G_4 \; \hat{\mathcal{T}}_4 (k) \; \tilde{u}(k,t)  
- \frac{1}{M} \mathcal{F} \{ F \left( u(x,t) \right) \}  = 0, 
\ee
where $\tilde{u}(k,t)={\cal L} \hat{u}(k,t)$, and
we use finite parameters $G_2$ and $G_4$, that are defined by (\ref{G2G4}).

The inverse Fourier transform of (\ref{Eq-k2}) is
\be \label{Eq-x}
\frac{\partial^2}{\partial t^2} u(x,t) -
G_2 \; \mathcal{T}_2(x) \; u(x,t) -
G_4 \; \mathcal{T}_4(x) \; u(x,t) -
\frac{1}{\rho} f \left( u(x,t) \right) = 0 ,
\ee
where the finite parameters $G_2$ and $G_4$ are defined by (\ref{G2G4}).
Using (\ref{Eq-x})
the operators $\mathcal{T}_2(x)$ and $\mathcal{T}_4(x)$ are defined by
\be \label{Tx0}
\mathcal{T}_2(x) = 
\mathcal{F}^{-1} \{ \hat{\mathcal{T}}_2 (k) \} = 
+ \frac{\partial^2}{\partial x^2} , \quad
\mathcal{T}_4(x) =
\mathcal{F}^{-1} \{ \hat{\mathcal{T}}_4 (k) \} = 
- \frac{\partial^4}{\partial x^4} .
\ee
Here, we have used the connection between the derivatives of 
the second and fourth orders and their Fourier transforms 
$k^2 \longleftrightarrow - \partial^2/\partial x^2$
and $k^4 \longleftrightarrow + \partial^4/\partial x^4$, 
and equations (\ref{T2T4}) . 

As a result, we obtain continuum equations (\ref{CME}). 
This ends the proof. \\


Proposition 2 illustrates the close relation between the discrete microstructure 
and the gradient non-local continuum. 
Let us consider special cases of the suggested model.

Lattice equations (\ref{C1}) have 
two parameters $g_2$ and $g_4$.
The corresponding equation (\ref{CME}) for the elastic continuum 
has two finite parameters $G_2$ and $G_4$.
If we use $g_2= 4 \, K$ and $g_4=0$, 
then $G_2= C^2_e = K \, a^2/M$, $G_4=0$, and 
we get equation (\ref{CME0}). 
If we assume that  $g_2= 4 \, K$ and $g_4 = - 4 \, K$,
then $G_2= C^2_e = K \, a^2/M$, $G_4= C^2_e \, a^2/48$ 
and we get the equation  
\be \label{eq6b}
\frac{\partial^2 u(x,t)}{\partial t^2} = C^2_e \, 
\frac{\partial^2 u(x,t)}{\partial x^2} + 
\frac{a^2 \, C^2_e}{12} \, 
\frac{\partial^4 u(x,t)}{\partial x^4} + \frac{1}{\rho} f(u), 
\ee
where $C_e=\sqrt{E/\rho}$ is the elastic bar velocity. 
Equation (\ref{eq6b}) can also be derived 
by the homogenization procedure \cite{M1968,RRG1995,MO1996}. 

In general, the coupling constants $g_2$ and $g_4$ are independent. 
Therefore the coupling constant $g_4$ 
may differ from the constant $g_2=4 \, K$.
If the relation of stress and displacement 
of the form $\varepsilon (x,t) = \partial u(x,t)/ \partial x$ is used, 
and the continuum equation (\ref{CME}) is expresses as
\[ \rho \, \frac{\partial^2 u(x,t)}{\partial t^2} = \frac{\partial \sigma(x,t)}{\partial x} + f(u), \]
where $\rho=M/(A \, a)$, 
then the constitutive relation can be represented by
\be
\sigma = E \, \left( \frac{g_2}{4 \, K} \, \varepsilon - \frac{g_4 \, a^2}{48 K} \, 
\frac{\partial^2 \varepsilon}{\partial x^2} \right) ,
\ee
where we use $E=K \, a/A$.
Therefore, using the correspondence principle, we will assume 
$g_2 = 4 \, K$. 
The second-gradient parameter $l$ is defined by the relation
\be 
l^2 = \frac{|g_4| \, a^2}{48 \, K} ,
\ee
where the sign in front of the factor 
$l^2$ in the constitutive relation 
is determined by the sign of the coupling constant $g_4$.
If the constant $g_4$ is positive then we get 
the second-gradient model with negative sign. 
As a result the second-gradient model with positive and 
negative signs 
\be \label{Minus-1}
\sigma =E \left( \varepsilon - \operatorname{sgn}(g_4) \, l^2 \, \frac{\partial^2 \varepsilon}{\partial x^2} \right) 
\ee
can be derived from a microstructure of lattice particles by suggested approach.
The proposed model as shown above uniquely leads to second-order
strain gradient terms that are preceded by the positive and negative signs.
It should be noted that positive value of coupling constant 
$g_4$ of lattice model can lead 
to effective stiffness coefficient of the next-nearest-neighbor interaction with non-convex elastic energy potentials 
in the effective discrete mass-spring system.
The strain gradients in continuum equation  
with the negative sign are equivalent to those derived 
from the positive-definite deformation energy density, 
and therefore these continuum models are stable. 
The lattice models with negative value of coupling 
constant $g_4$ of lattice model leads to  
the continuum equation with the positive sign 
in front of the parameter $l^2$. 
This continuum equation is unstable for wave numbers $k> 1/l^2$ 
The instability leads to an unbounded growth
of the response in time without external work.



\setcounter{figure}{1}
\begin{figure}[H]
\begin{minipage}[h]{0.47\linewidth}
\resizebox{12cm}{!}{\includegraphics[angle=-90]{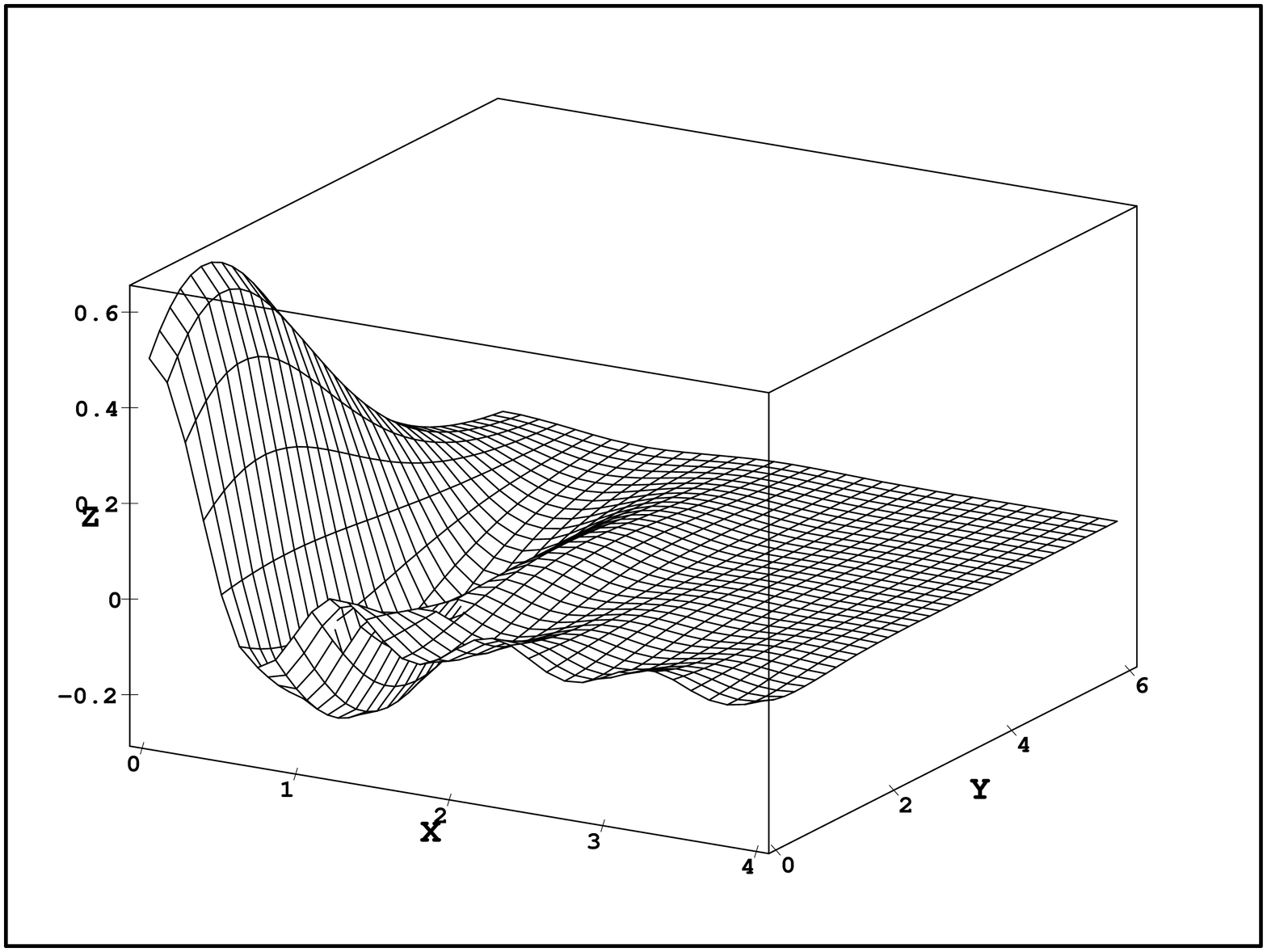}} 
\end{minipage}
\caption{Plot of the function (\ref{Knf}) 
for the range $x\in[0,4]$ and $y\in[0,6]$.} 
\label{Plot1}
\end{figure}


\begin{figure}[H]
\begin{minipage}[h]{0.47\linewidth}
\resizebox{12cm}{!}{\includegraphics[angle=-90]{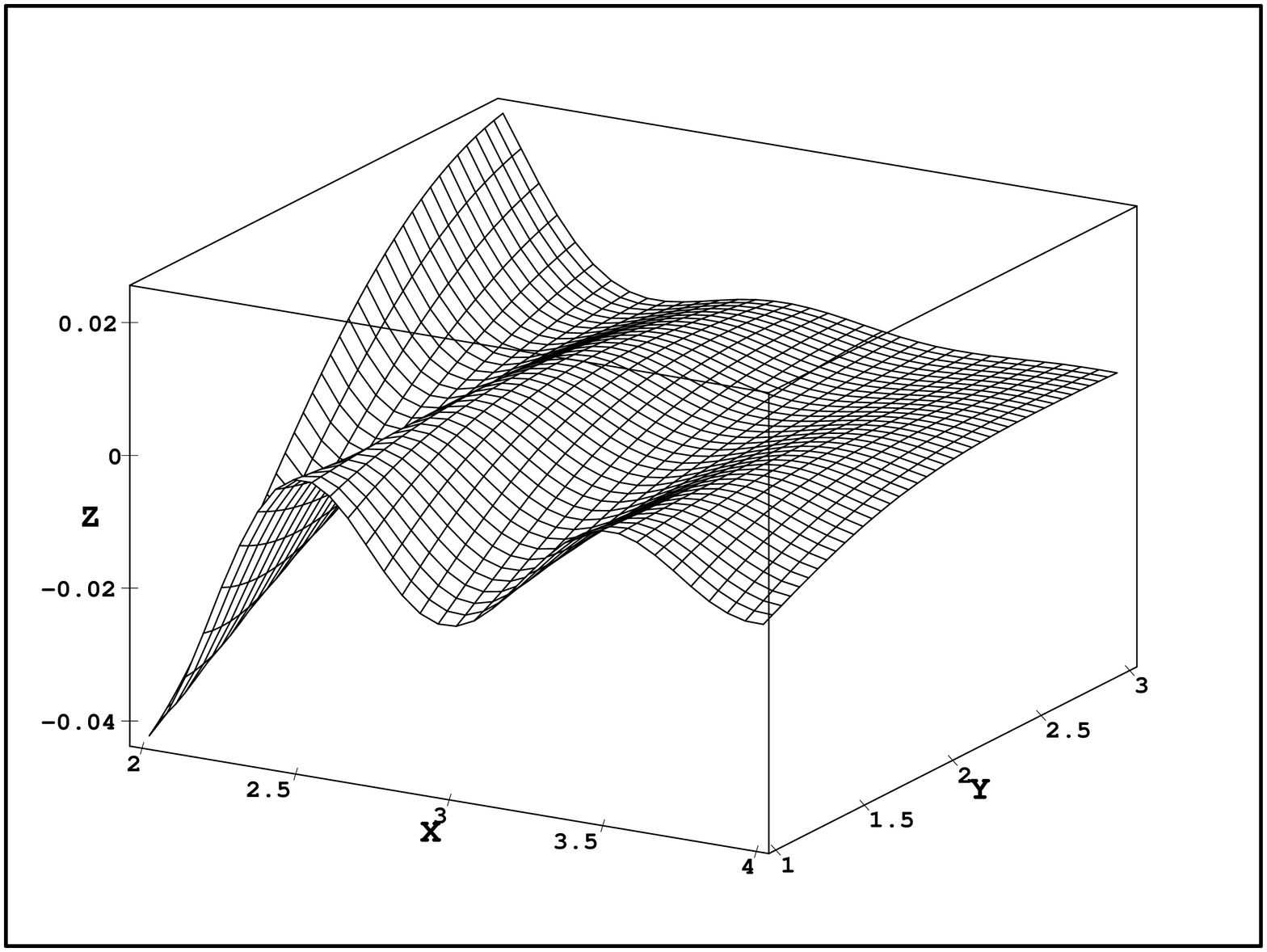}} 
\end{minipage}
\caption{Plot of the function (\ref{Knf}) 
for the range $x\in[2,4]$ and $y\in[1,3]$.} 
\label{Plot2}
\end{figure}

\section{Possible extensions of general lattice model}

The suggested lattice model can be generalized and 
extended for the high-order gradient elasticity and 
for three-dimensional lattice models.
Let us give some details about these generalizations. 


We can consider a generalization of the suggested lattice model 
by using the sum of the functions (\ref{Kn}) with 
the even value $s$.
Using the functions (\ref{Kn}) with $s=6$ and other even values,
we can consider the lattice models for high-order 
gradient elasticity \cite{ASS2002,AS2003,AM2005}. 
We can state that the lattice equations 
\be \label{C1h}
M \frac{d^2 u_n(t)}{d t^2} = 
\sum^N_{j=1} g_{2j} \sum_{\substack{m=-\infty \\ m \ne n}}^{+\infty} \, K_{2j}(|n-m|) \; \Bigl(u_n(t) -u_m(t)\Bigr) 
+ F (u_n) ,
\ee 
where $g_{2j}$ ($j=1,...,N$) are coupling constants,  
and $K_{2j}(|n-m|)$ are defined by (\ref{Kn}),
are transformed by the combination ${\cal F}^{-1} {\cal L} \ {\cal F}_{\Delta}$ 
of the operations (\ref{O1}-\ref{O3}) into the continuum equation 
\be \label{CMEh}
\frac{\partial^2 u(x,t)}{\partial t^2} + 
\sum^N_{j=1} (-1)^j \, G_{2j} \frac{\partial^{2j} u(x,t)}{\partial x^{2j} } 
- \frac{1}{\rho} f(u)= 0  ,
\ee
where $G_{2j} = g_{2j} \, a^{2j}/ (2\, \Gamma(2j+1))$,  
$(j=1,...,N)$ are finite parameters. 
The proof of this statement is similar to proof of Proposition 2. 


The suggested one-dimensional lattice model 
for second-gradient elasticity
can also be generalized for the three-dimensional case. 
We may consider a three-dimensional lattice
that is described by the equations
\be \label{2-3E1}
\frac{d^2 u^k_{\bf n} }{dt^2} =
\sum_{\substack{{\bf m}: {\bf m} \ne {\bf n}}} \;
K^{kl}_2({\bf n}-{\bf m} ) \; \left( u^l_{\bf n} - u^l_{\bf m} \right) +
\sum_{\substack{{\bf m}: {\bf m} \ne {\bf n}}} \;
K^{kl}_4 ({\bf n}-{\bf m} ) \; \left( u^l_{\bf n} - u^l_{\bf m} \right) 
+ F^k (u_{\bf n}) ,
\ee
where ${\bf n}=(n_1,n_2,n_3)$, $k,l=1,2,3$ and 
we assume a sum over repeated index $l=1,2,3$.
In the model (\ref{2-3E1}) the coupling constants 
are included in the tensors
$K^{kl}_s({\bf n}-{\bf m})=K^{kl}_s({\bf m}-{\bf n})$ 
that are distinguished by 
different power-law asymptotic behavior.
We assume that the functions  
$\hat{K}^{kl}_s({\bf k}) - \hat{K}^{kl}_s(0)$ 
are asymptotically equivalent to $k_i \, k_j$ 
and  $k_i \, k_j \, |{\bf k}|^2$ for $s=2$ and $s=4$ 
respectively, where
\[ \hat{K}^{kl}_s({\bf k})=
\sum_{{\bf n}} \; e^{-i {\bf k} \, {\bf n}} \; K^{kl}_s({\bf n}) . \]
To get continuum equation, we consider 
the field $u_{\bf n}(t)$ as Fourier coefficients
of the function $\hat{u}({\bf k},t)$, 
where ${\bf k}=(k_1,k_2,k_3)$, by
\[ \hat{u}^k({\bf k},t) = \sum_{{\bf n}} \;
u^k_{\bf n}(t) \; e^{-i {\bf k} {\bf r}_{\bf n}} , \]
where ${\bf r}({\bf n}) = {\bf r}_{\bf n}=
\sum^3_{i=1} n_i \, {\bf a}_i$, 
with the translational vectors ${\bf a}_i$ of the lattice.
In three-dimensional lattice model for 
second-gradient elasticity, we should consider 
interaction terms $K^{kl}_s  ({\bf n}-{\bf m} )$ 
that satisfy the conditions
\be \label{2-Aa2}
\lim_{k_i , k_j \to 0}
\frac{\hat{K}^{kl}_2({\bf k})- \hat{K}^{kl}_2(0)}{k_i \, k_j}
= A^{kl}_{ij}(2) , \quad
\lim_{k_i , k_j, |{\bf k}| \to 0}
\frac{\hat{K}^{kl}_4({\bf k})- \hat{K}^{kl}_4(0)}{k_i \, k_j \, 
|{\bf k}|^2} =A^{kl}_{ij}(4),
\quad (i,j=1,2,3) ,
\ee
where $A^{kl}_{ij}(s)$ are the coupling constants.
In the continuous limit ($|{\bf a}_i| \to 0$),
the three-dimensional lattice (\ref{2-3E1}) 
gives the continuum equations in the form
\be \label{2-3E1b}
\frac{\partial^2  u^k({\bf r},t) }{\partial t^2}
- G^{kl}_{ij}(2)
\frac{\partial^2 u^l({\bf r},t) }{\partial x_i \partial x_j}
+ G^{kl}_{ij}(4)
\frac{\partial^{4} u^l({\bf r},t) }{ \partial x_i \partial x_j
\partial x_m \, \partial x_m }
-\frac{1}{\rho} \, f^k(u({\bf r},t)) =0 , \ee
where we assume a sum over repeated indices $i,j,l,m \in\{1,2,3\}$, and
\be
G^{kl}_{ij}(2) = \frac{|{\bf a}_i| \, |{\bf a}_j| }{M} \, A^{kl}_{ij}(2) , \quad 
G^{kl}_{ij}(4) = \sum^3_{m=1} \frac{|{\bf a}_i| \, |{\bf a}_j| \, |{\bf a}_m|^2}{M} \, A^{kl}_{ij}(4) ,
\ee
where no summation over repeated indices.
We can consider the case with  
$G^{kl}_{ij}(4)=l^2 \, G^{kl}_{ij}(2)$, 
where $G^{kl}_{ij}(2)=C_{ikjl}$ can be considered
as a stiffness tensor and $l^2$ is the scale parameter. 
For isotropic case, we have 
$C_{ijkl}= \lambda \, \delta_{ij} \delta_{kl} +
\mu \, (\delta_{ik} \delta_{jl} +\delta_{il} \delta_{jk} )$.
In general equation (\ref{2-3E1b}) describes 
anisotropic gradient continuum.
A more detailed description of the three-dimensional lattice
model (\ref{2-3E1}) will be made in the following article.

\section{Conclusion}

In this paper lattice models 
for strain-gradient elasticity of continuum are suggested.
The first advantage of the suggested lattice models 
is a possibility to consider these models 
as a microstructural basis of unified description 
of gradient models with positive and negative signs 
of the strain gradient terms.
A second advantage of the proposed model is that 
it can be easily generalized to the case of 
the high-order gradient elasticity by using the 
other even values of $s$. 
Using (\ref{Kn}) with positive integer $s=2j$,
we have non-local interaction of the lattice particle, 
that gives the derivatives of integer order $2 j$  
in the continuum equation. 
Three-dimensional lattice models 
and the correspondent continuum equation can also be formulated 
as (\ref{2-3E1}) and (\ref{2-3E1b}).
The third advantage of the proposed form of the interaction 
is that the lattice equations can be used not only 
for the integer but also for fractional values 
of the parameter $s$. 
Therefore the suggested general lattice model can be 
extended on the fractional nonlocal case.
The suggested types (\ref{Kn}) and (\ref{Kn2}) of inter-particle 
interactions in the lattice can be used for non-integer $s$.
If we consider interaction terms defined by (\ref{Kn}) 
and (\ref{Kn2}) with non-integer $s=\alpha$, 
then we will get continuum equations with the Riesz 
fractional derivatives \cite{KST} of orders $s=\alpha$ 
by the methods suggested in \cite{JMP2006,JPA2006}.
The lattice models with long-range interactions
of the types (\ref{Kn}) and (\ref{Kn2}) 
with non-integer $s=\alpha$,
can serve as microscopic models for elastic continuum 
with power-law non-locality. 
It allows us to derive fractional generalizations 
of gradient elasticity by using 
a microscopic approach \cite{CEPJ2013,MOM2014}.
We also assume that the suggested lattice model can be
generalized to get discrete (lattice) models for 
dislocations in the gradient elasticity continuum
\cite{GA-1,GA-2,Lazar1,Lazar2,Lazar2b,Lazar3,Lazar4,Lazar5},
and then it will be possible to extend them 
to the fractional non-local case. 
The suggested lattice models with long-range interactions,
which are suggested for the gradient elasticity continuum,
can be important to describe the non-local elasticity 
of materials 
at micro and nano scales \cite{MOS2002,Nano,Nano1,Nano2},
where the interatomic and intermolecular interactions 
are prevalent in determining the properties of these materials.



\end{document}